\documentclass[smallextended]{svjour3} 

\smartqed  

\usepackage{amsmath,amssymb}
\usepackage{chngcntr}
\usepackage{mathtools}
\usepackage{graphicx}
\usepackage{color}
\usepackage[all]{xy}

\usepackage{theorem}

{\theorembodyfont{\upshape}
}

{\theorembodyfont{\upshape}
}
{\theorembodyfont{\upshape}
}
{\theorembodyfont{\upshape}
}

\numberwithin{thm}{section}
\numberwithin{equation}{section}

\newcommand{\bE}{\mathbb{E}}

\newcommand{\bN}{\mathbb{N}}

\newcommand{\bP}{\mathbb{P}}
\newcommand{\bQ}{\mathbb{Q}}
\newcommand{\bR}{\mathbb{R}}

\newcommand{\calI}{\mathcal{I}}
\newcommand{\calJ}{\mathcal{J}}

\newcommand{\calL}{\mathcal{L}}

\newcommand{\calN}{\mathcal{N}}

\newcommand{\PZ}{\mathbb{Z}_{+}}

\journalname{Journal of Statistical Physics}

\begin{document}

\title{Macroscopic dynamical fluctuations in Kac ring model}

\author{Ken Hiura}

\institute{K. Hiura \at
              Department of Physics, Kyoto University, Kyoto 606-8502, Japan \\
              \email{hiura.ken.88n@st.kyoto-u.ac.jp} }

\date{Received: date / Accepted: date}

\maketitle

\begin{abstract}
We study dynamical fluctuations in the macroscopic paths around the most probable path of the Kac ring model, which is a simple deterministic and reversible dynamical system exhibiting the macroscopic irreversible relaxation. We derive the form of the generating function for macroscopic paths and show that small deviations are described by a discrete-time Ornstein-Uhlenbeck process. We also argue that the microscopic reversibility leads to the fluctuation relation of the rate function, and prove it based on the form of the generating function.

\end{abstract}

\section{Introduction}

A macroscopic system in equilibrium consists of an enormous number of microscopic constituents and its microscopic dynamical behavior is strongly chaotic. Therefore, the dynamical properties at small scales are unpredictable. However, the instantaneous macroscopic observations for microscopically different states of the system provide the same outcome, which is described by its thermodynamic function. This reproducibility of thermodynamic observations is guaranteed by the law of large numbers. Surprisingly, as well as the most probable values of thermodynamic quantities, even the statistics of the fluctuations around these values are also described by the thermodynamic function. This theory of fluctuations was developed by Einstein \cite{Einstein1910} and is regarded as the large deviation theory for static fluctuations in equilibrium systems \cite{Lanford1973,Pfister2002}. Moreover, if we focus on small deviations from the most probable value, we find a universal Gaussian structure of the fluctuations that is accordance with the central limit theorem \cite{MartinLof1979}.

Extending the theory for static fluctuations to dynamical fluctuations has been an important topic in nonequilibrium statistical physics. Pioneering work by Onsager and Machlup developed the linear theory for dynamical fluctuations in equilibrium \cite{OnsagerMachlup1953,MachlupOnsager1953}. This framework was later applied to nonequilibrium processes, e.g., \cite{BertiniDeSoleGabrielliJonaLasinioLandim2007}. The method by Onsager and Machlup based on stochastic models is also useful in studying the long-term behavior of Markov processes and leads to an important class of results in nonequilibrium statistical mechanics, namely the fluctuation relations, which are the symmetry properties of the fluctuations of the entropy production \cite{EvansCohenMorris1993,GallavottiCohen1995a,GallavottiCohen1995b,Kurchan1998,LebowitzSpohn1999,Maes1999}. Nevertheless, we do not have a complete understanding of the \textit{derivation} of the large deviation property and stochastic models describing fluctuations from the underlying microscopic deterministic dynamics. For instance, despite the fluctuating hydrodynamics being a useful tool in analyzing the long-wavelength and long-time fluctuations \cite{LandauLifshitz1958,OrtizdeZarateSengers2006,ForsterNelsonStephen1977}, its derivation from Hamiltonian dynamics is still at a formal level \cite{ZubarevMorozov1983}. Moreover, the numerical verification for anharmonic chains has only been studied recently \cite{SaitoHongoDharSasa2021}. We note, however, that there was an important recent mathematical development concerning the theory of fluctuations for dilute gases in which the constituent particles obeyed Newtonian dynamics with the hard-core interaction \cite{BodineauGallagherSaintRaymondSimonella2020}.

The purpose of this paper is to illustrate a route from a microscopic reversible dynamical system to its macroscopic dynamical fluctuations through a simple toy model. Specifically, we study the dynamical fluctuations of the Kac ring model, which was introduced by Mark Kac \cite{Kac1959} as a toy model demonstrating how macroscopic irreversibility is obtained from a microscopically reversible dynamical system. He showed that despite the Kac ring model being microscopically reversible, its macroscopic observable exhibits the relaxation behavior in the thermodynamic limit. His result corresponds to the law of large numbers associated with the most probable macroscopic path. We shall here investigate the large deviation property for fluctuations around the most probable macroscopic path. The two main contributions of this paper are deriving the form of the generating function for the macroscopic path of the Kac ring model and finding the Gaussian stochastic process that describes the small fluctuations. Although the Kac ring model is not a realistic model of some phenomenon in nature, it allows exact calculations and we believe that it is instructive for our purposes.

The remainder of the paper is organized as follows. In Section \ref{sec:model}, we introduce the Kac ring model and review its basic features, in particular, the law of large numbers. In Section \ref{sec:fluctuation}, we derive the form of the generating function and study small fluctuations. In Section \ref{sec:fr}, we argue that the microscopic reversibility leads to a fluctuation relation of the rate function and give a proof based on the form of the generating function. Concluding remarks are given in Section \ref{sec:conclusion}.


\section{Model and Its Properties}
\label{sec:model}

We begin by reviewing the basic properties of the Kac ring model. See also \cite{Kac1959,GottwaldOliver2009,MaesNetocnyShergelashvili2009}. Throughout the paper, we use the notation $\bP [ \{ x \in X : C(x) \} ] \eqqcolon \bP [ C ]$ for a probability measure $\bP$ on a space $X$ and a condition $C$.

\subsection{Model}
\label{subsec:model}

\begin{figure}
\centering
\includegraphics[width=11cm]{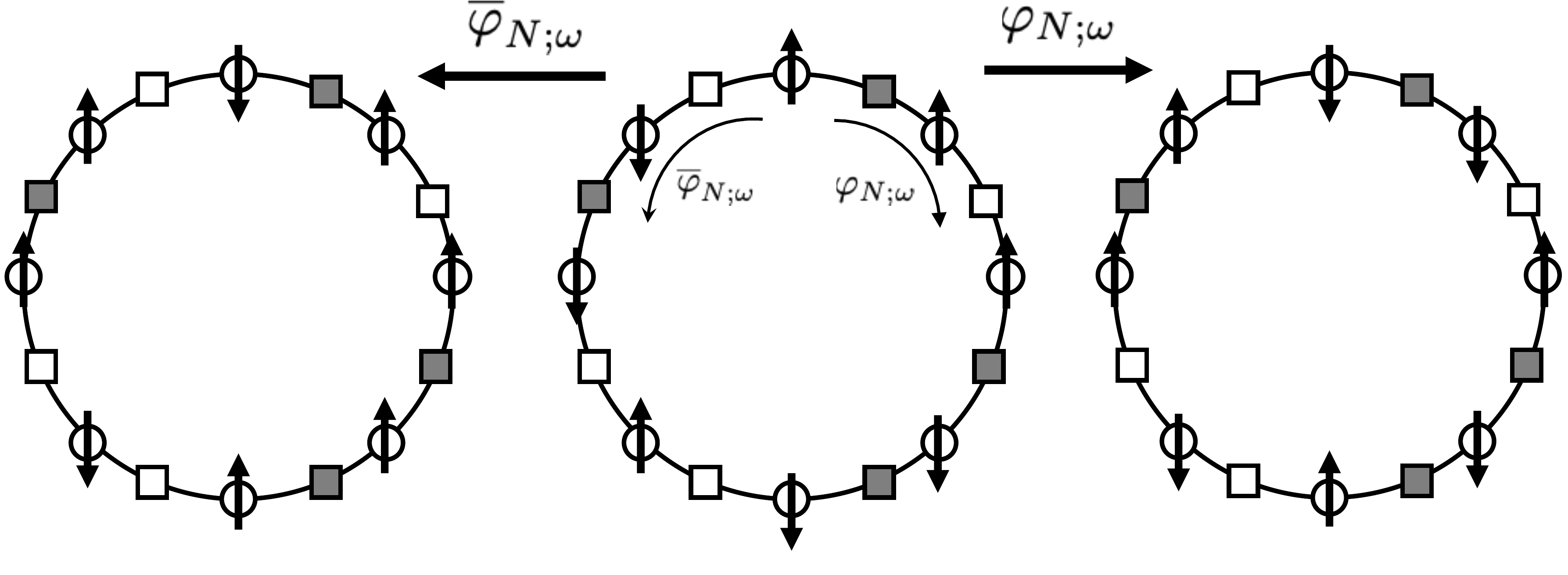}
\caption{Schematic of the time evolution of the Kac ring model. On the sites, up and down arrows in circles represent spin up and spin down, respectively. Gray boxes between circles represent scatterers.}
\label{fig:kacring}
\end{figure}

We consider a ring $T_N = \{ 1, 2, \dots, N \}$ of the length $N \in \PZ$ and impose the periodic boundary condition, i.e., $N+1 = 1$. On each site $i \in T_N$, we have a spin variable $\sigma_N (i) \in S \coloneqq \{ -1, 1 \}$ and, as quenched disorder, an occupation number variable of a scatterer $\omega_N (i) \in B = \{ 0, 1 \}$. Therefore, the phase space is given by $S^{T_N}$ with the space of quenched variables $B^{T_N}$. For a one-sided infinite sequence on $B$, $\omega = ( \omega(1), \omega(2), \dots ) \in B^{\PZ}$, we use the same notation $\omega_N = ( \omega(1), \omega(2),  \dots,  \omega(N)) \in B^{T_N}$ to denote the restriction of $\omega$ onto the first $N$ bits. The Kac ring model is a discrete-time dynamical system on the phase space $S^{T_N}$. The transformation map $\varphi_{N; \omega} : S^{T_N} \to S^{T_N}$ is given by
\begin{align} \label{eq:map}
 \varphi_{N; \omega} (\sigma_N)(i) = ( 1 - 2 \omega_N(i-1)) \sigma_N (i-1), \ \ \sigma \in S^{T_N}, \ i \in T_N
\end{align}
for $\omega \in B^{\PZ}$. The intuitive meaning of the map $\varphi_{N;\omega}$ is as follows. We consider a spin $\sigma_N (i)$ on a site $i \in T_N$ at time $t \in \bN$. If a scatterer is present on the same site $i$, i.e., $\omega(i) = 1$, the spin jumps to its neighboring site $i+1$ and its direction flips. If no scatterer is on site $i$, i.e., $\omega(i) = 0$, the spin jumps to site $i+1$ but its direction remains unchanged (see Fig. \ref{fig:kacring}). The variable $\omega$ signifies quenched disorder because its value does not change as time elapses.

The Kac ring model shares two important properties with the Hamiltonian dynamics. First, the model preserves the phase space volume, i.e, the counting measure: $| \varphi_{N;\omega}^{-1} (A)| = |A|$ for any subset $A \subseteq S^{T_N}$. From this property, we find that the uniform distribution $U_N$ on $S^{T_N}$ is an invariant measure of the map $\varphi_{N;\omega}$ for any $\omega$, and that it is regarded as the equilibrium state of the Kac ring model. Second, the Kac ring model is reversible. Indeed, the map
\begin{align} \label{eq:inversemap}
 \overline{\varphi}_{N;\omega}(\sigma_N)(i) = ( 1 - 2 \omega_N (i)) \sigma_N (i+1), \ \ \sigma \in S^{T_N}, \ i \in T_N
\end{align}
is the inverse of $\varphi_{N;\omega}$. We discuss the reversibility in more detail in Section \ref{sec:fr}.

We remark that any configuration $\sigma_N$ of spins are periodic points of $\varphi_{N;\omega}$ and the period is less than $2N$. This follows from $\varphi_{N;\omega}^{2N}(\sigma_N) = \sigma_N$. We easily understand this periodicity from the fact that each spin experiences flipping twice at each site on which a scatterer sits when the ring has been rotated twice.

\subsection{Law of Large Numbers}

The Kac ring model exhibits macroscopic relaxation behavior, while being microscopically reversible. To see this, we introduce two macroscopic variables, the magnetization $m_N$:
\begin{align}
 m_N (\sigma_N) \coloneqq \frac{1}{N} \sum_{i=1}^{N} \sigma_N (i).
\end{align}
If we regard the uniform distribution $U_N$ as the equilibrium distribution of the spins, the equilibrium value of the magnetization is given by $m_{\mathrm{eq}} \coloneqq \bE_{\sigma_N \sim U_N} [m_N] =  0$. We now suppose that we have an initial macroscopic state specified by the value of the magnetization, $m_N \approx m$, in the scatterers induced by $\omega = (\omega(i))_{i \in \PZ} \in B^{\PZ}$. We expect that if the initial magnetization is away from equilibrium, i.e., $m \neq 0$, the magnetization undergoes relaxation in time as a consequence of the scattering induced by the quenched disorder. As a probability distribution on the phase space $S^{T_N}$ describing the macroscopic state specified by $m_N \approx m$, we introduce canonical measure,
\begin{align} \label{eq:cm}
 \bP_{N;m} [ \sigma_N ] = \exp \left[ N \left( \beta m_N (\sigma_N)- P(\beta) \right) \right] U_N [ \{ \sigma_N \} ],
\end{align}
where $P(\beta) = \ln (\cosh (\beta))$ denotes the free energy, and parameter $\beta$ is determined from condition $\bE_{N;m} [ m_N ] = m$. Here, $\bE_{N;m}$ denotes the expectation value with respect to $\bP_{N;m}$. This equation is easily solved, giving $\beta = \frac{1}{2} \ln (1+m)/(1-m)$, and thus we obtain that
\begin{align}
 \forall i \in T_N, \ \bP_{N;m} [ \sigma_N(i) = \pm 1 ] = \frac{1 \pm m}{2}.
\end{align}
Because the dynamical behavior of the magnetization depends on the configuration of the scatterers $\omega$, we introduce a probability distribution on the scatterers. For this distribution, we choose the Bernoulli distribution denoted here by $\bQ_{\rho}$ with parameter $\rho \in (0,1)$. This implies that, independent of all other sites, a scatterer is present at site $i$ with probability $\rho$. Similarly to the choice of the initial distribution (\ref{eq:cm}) for the spins, this distribution is also regarded as the canonical measure determined from condition $\bE [ \omega (i) ] = \rho$ for a given density of scatterers $\rho$. We note that there are two settings, quenched and annealed. In the quenched setting, we perform experiments with a fixed configuration of scatterers and study the statistical property of $\bE_{N;m} [ f(\cdot, \omega) ]$ with respect to the distribution $\bQ_{\rho}$, where $f(\sigma_N, \omega)$ is an observable. In the annealed setting, we prepare a configuration of scatterers every time we perform an experiment. That is, we investigate the expectation $f \mapsto \int \bE_{N;m} [ f(\cdot, \omega)] \bQ_{\rho}(d\omega)$ in the annealed setting. In this study, we mainly focus on the quenched setting; we shall only make a comment on the annealed setting at the end of Section \ref{sec:fluctuation}. We remark that both for both settings, the configuration $\omega$ does not change with time and therefore it is regarded as \textit{quenched disorder in the time direction} in a single experiment.

The statistical behavior of the time evolution for magnetization is determined by these probability distributions. If we define the empirical magnetization at time $t$ as
\begin{align}
 \mu_{N, t} (\sigma_N, \omega) \coloneqq m_N ( \varphi_{N;\omega}^t (\sigma_N)),
\end{align}
we easily find that
\begin{align}
 \lim_{N \to \infty} \bE_{N;m} [ \mu_{N,t} (\cdot, \omega) ] = ( 1 - 2 \rho)^t m \eqqcolon \Phi_{\rho}^t (m)
\end{align}
for $\bQ_{\rho}$-almost every $\omega$, which implies that the magnetization relaxes exponentially in time to the equilibrium value $m_{\mathrm{eq}} = 0$ on average, and the relaxation behavior is characterized by the map $\Phi_{\rho} (m) = (1 - 2\rho) m$. The macroscopic law $\Phi_{\rho}$ describes the typical behavior of the magnetization as well as the average behavior. This property is formulated as a law of large numbers: for any small positive real number $\epsilon > 0$ and any time $T \in \bN$, 
\begin{align} \label{eq:qlln}
 \lim_{N \to \infty} \bP_{N;m} \left[ \left( \exists t \in [0,T] \right) \left( |\mu_{N,t} (\cdot, \omega) - \Phi_{\rho}^t (m) | > \epsilon \right) \right] = 0, \ \ \bQ_{\rho}\text{-a.s.}
\end{align}
Although the original analysis by Kac concerned the annealed setting with Eq. (\ref{eq:alln}), it is straightforward to prove the law of large numbers in the quenched setting. The law of large numbers implies that the probability of the empirical macroscopic path $(\mu_{N,t})_{t \in [0,T]}$ concentrates on path $(\Phi_{\rho}^t (m))_{t \in [0,T]}$ for large systems. Therefore, even if we perform the experiment only once, we observe the magnetization relaxation described by the macroscopic law $\Phi_{\rho}$ with high probability when $N \gg 1$. Throughout this paper, we first fix a time interval $[0,T]$ and then take the limit $N \to \infty$ as in Eq. (\ref{eq:qlln}). Therefore, the recurrence property of the underlying dynamical system does not matter.

\section{Dynamical Fluctuations in Kac ring model}
\label{sec:fluctuation}

In this section, we derive the form of the generating function for the magnetization in the quenched setting and find the formula relating it to the generating function in the annealed setting. From that result, we specify the discrete-time Ornstein-Uhlenbeck process that governs small deviations in the quenched setting and obtain the Onsager-Machlup form for the path probability density.

\subsection{Large Deviations}
\label{subsec:ld}

In Eq.(\ref{eq:qlln}), we now assume that the probability that the empirical path $(\mu_{N,t})_{t \in [0,T]})$ deviates from the most probable path $(\Phi_{\rho}^t(m))_{t \in [0,T]}$ is exponentially small with respect to system size $N$. This assumption is expressed in the large deviation form,
\begin{align} \label{eq:qldp}
 \bP_{N;m} \left[ \mu_{N,t} (\cdot, \omega) \approx m_t, \ t \in [0,T] \right] \asymp \exp \left( - N \calI^q_{m, \rho, [0,T]} (m_{[0,T]}) \right), \ \ \bQ_{\rho}\text{-a.s.},
\end{align}
where we use the notation $m_{[0,T]} = (m_t)_{t \in [0,T]}$ and interpret $a_n \asymp b_n$ as meaning that $n^{-1} \ln (a_n / b_n) \to 0$ as $n \to \infty$. We remark that the assertion (\ref{eq:qldp}) is expected to be true for almost all configurations $\omega \in B^{\PZ}$ with respect to $\bQ_{\rho}$. The rate function $\calI^q_{m,\rho,[0,T]}$ is called the \textit{quenched rate function}. We note that this rate functions must be non-negative and has a unique zero at $m_t = \Phi_{\rho}^t(m)$ for $t \in [0,T]$ in accordance with the laws of large numbers (\ref{eq:qlln}).

The question here is how to calculate the rate function. According to the G\"artner-Ellis theorem \cite{DemboZeitouni1998}, the quenched generating function defined by
\begin{align}
 \Lambda_{m,\rho,[0,T]}^q (k_{[0,T]}) \coloneqq \lim_{N \to \infty} \frac{1}{N} \ln \bE_{N;m} \left[ \exp \left( N \sum_{t=0}^{T} k_t \mu_{N,t} ( \cdot, \omega) \right) \right]
\end{align}
exists for $k_{[0,T]} = (k_t)_{t \in [0,T]} \in \bR^{T+1}$ and is differentiable in $k_{[0,T]}$ for $\bQ_{\rho}$-almost all $\omega$, the large deviation property (\ref{eq:qldp}) is then assured, and the rate function is given by the Legendre transform of $\Lambda_{m,\rho}^q$,
\begin{align} \label{eq:lt}
 \calI_{m,\rho,[0,T]}^q (m_{[0,T]}) = \sup_{k_{[0,T]} \in \bR^{T+1}} \left( \sum_{t=0}^{T} k_t m_t - \Lambda_{m,\rho,[0,T]}^q (k_{[0,T]}) \right).
\end{align}
Therefore, the problem is reduced to the calculation of the generating function. In the quenched setting, the generating function is easily obtained. We find that
\begin{align}
 &\frac{1}{N} \ln \bE_{N;m} \left[ \exp \left( N \sum_{t=0}^{T} k_t \mu_{N,t} ( \cdot, \omega) \right) \right] 
 \notag \\
 &= \frac{1}{N} \sum_{i=1}^{N} \ln \left[ \cosh \left( k_0 + \sum_{t=1}^{T} k_t (1-2 \omega_N(i)) \dots (1-2\omega_N(i+t-1)) \right) \right.
 \notag \\
 & \qquad \left. + m \sinh \left( k_0 + \sum_{t=1}^{T} k_t (1-2 \omega_N(i)) \dots (1-2\omega_N(i+t-1)) \right) \right].
\end{align}
The summands are correlated but identically distributed random variables because of translation invariance. Because the correlation length is at most $T$, the strong law of large numbers for the above empirical mean holds and therefore, in the limit $N \to \infty$, the empirical mean is replaced by the expectation of the summand with probability one. Therefore, we conclude that the quenched generating function exists and is given by
\begin{align} \label{eq:qgf}
 \Lambda_{m,\rho,[0,T]}^q (k_{[0,T]}) &= \sum_{y_1, \dots, y_T \in B} \rho^{\sum_{i=1}^{T} y_i} (1-\rho)^{T - \sum_{i=1}^{T} y_i}
 \notag \\
 & \qquad \ln \left[ \cosh \left( k_0 + \sum_{t=1}^{T} k_t (1-2 y_1) \dots (1-2y_t) \right) \right.
 \notag \\
 & \qquad \left. + m \sinh \left( k_0 + \sum_{t=1}^{T} k_t (1-2 y_1) \dots (1-2 y_t) \right) \right]
\end{align}
for almost every $\omega$ with respect to $\bQ_{\rho}$. In particular, the generating function for magnetization at time $t$ is given by
\begin{align} \label{eq:qgft}
 \Lambda_{m,\rho,t}^q (k_t) &\coloneqq \lim_{N \to \infty} \frac{1}{N} \ln \bE_{N;m} \left[ \exp \left( N k_t \mu_{N,t}(\cdot, \omega) \right) \right] 
 \notag \\
 &= \frac{1 - (1-2\rho)^t}{2} \ln ( \cosh k_t - m \sinh k_t)
 \notag \\
 & \quad + \frac{1 + (1-2\rho)^t}{2} \ln (\cosh k_t + m \sinh k_t).
\end{align}
Eq. (\ref{eq:qgf}) is the first main result of this paper. The quenched generating function $\calI_{m,\rho,[0,T]}^q$ is obtained from Eq. (\ref{eq:qgf}) through the Legendre transformation (\ref{eq:lt}). Although the explicit form of the quenched rate function for general $T$ is difficult to express, we present explicit forms of $\calI_{m,\rho,[0,T]}^q$ for $T = 0$ and $T=1$ in Appendix \ref{app:explicit}. The form of the generating function (\ref{eq:qgft}) implies a peculiar property that although the magnetization decays to the equilibrium value and the averaged value loses the information on the initial value $m$ in the long time limit $t \to \infty$, this information ever remains in the fluctuations.

\subsection{Central Limit Theorems}
\label{subsec:clt}

We next consider small deviations of order $O(N^{-1/2})$ around the most probable path, which is described by the probability distribution of the normalized deviation,
\begin{align}
 \xi_{N,t} (\sigma_N, \omega) \coloneqq \sqrt{N} ( \mu_{N,t} ( \sigma_N, \omega ) - \Phi_{\rho}^t (m)).
\end{align}
More precisely, our purpose is to find the discrete-time Gaussian process to which $\xi_{N,t}$ converges in law on $[0,T]$ in the limit $N \to \infty$. This problem is called the \textit{central limit theorem}. In terms of the rate function, the probability distribution of the Gaussian process to which the normalized deviation converges in law is obtained from the second-order expansion of the rate function around the unique zero. From Eq. (\ref{eq:lt}), the covariance matrix characterizing the Gaussian process is determined from the Hessian matrix of the generating function at $k_{[0,T]}=0$ \cite{Bryc1993}. In the quenched setting, we have from Eq. (\ref{eq:qgf}) that
\begin{align} \label{eq:qcov}
 \left. \frac{\partial}{\partial k_s} \frac{\partial}{\partial k_t} \Lambda_{m,\rho,[0,T]}^q (k_{[0,T]}) \right|_{k_{[0,T]}=0} = (1 - 2 \rho)^{|t-s|} (1 - m^2).
\end{align}
We now consider the discrete-time Ornstein-Uhlenbeck process $\xi_t^q$ that obeys the stochastic difference equation,
\begin{align} \label{eq:qOU}
 \xi_{t+1}^q = \Phi_{\rho} (\xi_t^q) + \sqrt{2 D_{m,\rho}}  w_t, \ \ t \in \bN,
\end{align}
with the initial data $\xi_0^q \sim \calN (0, 1-m^2)$. Here, $D_{m,\rho} = 2 \rho (1-\rho) (1-m^2)$ is the diffusion constant, $\{ w_t \}_{ t \in \bN}$ is the i.i.d. sequence of Gaussian random variables of zero mean and variance one, and they are independent of $\xi_0^q$. Since the joint distribution of $(\xi_0, \dots, \xi_T)$ is normal for any $T$, the process $(\xi_t^q)$ is a Gaussian process. Moreover, we find that the process $(\xi_t^q)$ has the same covariance matrix as (\ref{eq:qcov}),
\begin{align} 
 \bE \xi_t^q \xi_s^q = (1- 2 \rho)^{|t-s|} \bE \left[ (\xi_{0}^q)^2 \right] = (1 - 2 \rho)^{|t-s|} (1-m^2).
\end{align}
Therefore, in the quenched setting, the path of the normalized deviation $\{ \xi_{N,t} ( \cdot, \omega) \}$ converges on $[0,T]$ in law to $\{ \xi_t^q \}$ for $\bQ_{\rho}$-almost every $\omega$. This central limit theorem is the second main result of the paper.

The central limit theorem is expressible in large deviation form. From the covariance matrix (\ref{eq:qcov}), if we focus on small deviations, the deviation probabilities in the quenched setting are given approximately by the Onsager-Machlup form \cite{OnsagerMachlup1953,MachlupOnsager1953},
\begin{align} \label{eq:om}
 \bP_{N;m} \left[ \mu_{N,t}(\cdot, \omega) \approx m_t, t \in [0,T] \right] \asymp  \exp \left( - N \left[ \calL_m^{\mathrm{ini}} (m_0) + \sum_{t=0}^{T-1} \calL_{m,\rho}^q (\dot{m}_t, m_t) \right] \right),
\end{align}
where $\dot{m}_t \coloneqq m_{t+1} - m_t$ is a velocity-like variable, $\calL_m^{\mathrm{ini}}(m_0) \coloneqq (m_0 - m)^2 / (1-m^2)$ is a quadratic term in the rate function at the initial time, and
\begin{align}
 \calL_{m,\rho}^q (\dot{m}_t, m_t) \coloneqq \frac{(\dot{m}_t + 2 \rho m_t )^2}{4 D_{m,\rho}}
\end{align}
is the quadratic Lagrangian that describes the transition from $m_t$ to $m_{t+1} = m_t + \dot{m}_t$.

We comment on the need for temporal coarse-graining. Generically, the separation of time scales and temporal coarse-graining play an essential role when we construct the stochastic models describing the dynamic fluctuations. Nevertheless, in the Kac ring model, we have the large deviation property and the stochastic model for small deviations at the microscopic time scales. This special property stems from the fact that the law of large numbers for the macroscopic path holds without the temporal coarse-graining [see Eq. (\ref{eq:qlln})]. In general, temporal coarse-graining permits the macroscopic law to be obtained and, correspondingly, we only expect the large deviation property and the central limit theorem when we observe the system at the appropriate time scale.

\subsection{Annealed Setting}

We make several comments on the annealed setting. We obtain the annealed distribution $\bP_{N;m,\rho}$ of the spins and the corresponding expectation $\bE_{N;m,\rho}$ by averaging the quenched distributions over all configurations of scatterers with respect to $\bQ_{\rho}$, i.e., $\bP_{N;m,\rho} ( \cdot ) = \int \bP_{N;m} ( \cdot ) \bQ_{\rho} (d \omega)$. Whenever confusion occurs, we use the same notation $\bP_{N;m,\rho}$ and $\bE_{N;m,\rho}$ to denote, respectively, the joint probability distribution of the spins and scatterers and the corresponding expectation. 

Similar to the quenched setting, we have the weak law of large numbers in the annealed setting \cite{MaesNetocnyShergelashvili2009},
\begin{align} \label{eq:alln}
 \lim_{N \to \infty} \bP_{N;m,\rho} \left[ \left( \exists t \in [0,T] \right) \left( |\mu_{N,t} - \Phi_{\rho}^t (m) | > \epsilon \right) \right] = 0
\end{align}
for any positive small real number $\epsilon > 0$ and any time $T \in \bN$. The large deviation property in the annealed setting takes the form
\begin{align} \label{eq:aldp}
 \bP_{N;m,\rho} \left[ \mu_{N,t}  \approx m_t, \ t \in [0,T] \right] \asymp \exp \left( - N \calI^a_{m, \rho,[0,T]} (m_{[0,T]}) \right).
\end{align}
The corresponding rate function $\calI_{m,\rho}^a$ is called the \textit{annealed rate function}, which is related through the Legendre transformation to the annealed generating function defined by
\begin{align} \label{eq:agf}
 \Lambda_{m,\rho,[0,T]}^a (k_{[0,T]}) \coloneqq \lim_{N \to \infty} \frac{1}{N} \ln \bE_{N;m,\rho} \left[ \exp \left( N \sum_{t=0}^{T} k_t \mu_{N,t} \right) \right].
\end{align}
The concavity of the logarithm implies that $\Lambda_{m,\rho,[0,T]}^a \geq \Lambda_{m,\rho,[0,T]}^q$ and therefore $\calI_{m,\rho,[0,T]}^a \leq \calI_{m,\rho,[0,T]}^q$. Intuitively, this inequality is understood as follows. Atypical fluctuations of the scatterers are allowed in the annealed setting, and hence large deviation probabilities in the annealed setting are not smaller than those in the quenched setting. 

In contrast to the quenched setting, calculating the limit (\ref{eq:agf}) in the annealed generating function is difficult. However, it is possible to calculate exactly the covariance matrix, and hence we obtain the central limit theorem in the annealed setting if we assume the existence and the sufficient analyticity of the annealed rate function. We find that in the annealed setting, the covariance matrix has a slightly complicated form because of the fluctuation in density of the scatterers,
\begin{align} \label{eq:acov}
  &\left. \frac{\partial}{\partial k_s} \frac{\partial}{\partial k_t} \Lambda_{m,\rho,[0,T]}^a (k_{[0,T]}) \right|_{k_{[0,T]}=0} 
  \notag \\
  &= (1-2\rho)^{|t-s|} \left[ \frac{1 - (1 - 2m^2)(1-2\rho)^2}{1 - (1-2\rho)^2} + |t-s|m^2  \right]
  \notag \\
  & \quad - \left[ \frac{1 + (1-2\rho)^2}{1-(1-2\rho)^2} + t+s \right] (1-2\rho)^{t+s} m^2,
\end{align}
for $t, s \geq 0$ and the corresponding Gaussian process is non-Markovian as long as the initial state is prepared to be out of equilibrium, i.e., $m \neq 0$.

\section{Microscopic Reversibility and Fluctuation Relation}
\label{sec:fr}

In this section, we study the fluctuation symmetry of the rate function. Although we analyze it in the quenched setting, a similar analysis is possible in the annealed setting.

\subsection{Microcanonical Setup}

We have considered the canonical setup, where the initial configurations are prepared according to the canonical measure (\ref{eq:cm}). Below, we introduce the rate function in the microcanonical setup to elucidate the fluctuation symmetry that arises from microscopic reversibility. The microcanonical measure is defined as the uniform distribution $U_N$ conditioned on the initial magnetization $m_N \approx m_0$, $U_N [ \cdot | m_N (\cdot) \approx m_0]$. We now suppose the large deviation property in the microcanonical setup,
\begin{align} \label{eq:mcld}
 U_N \left[ \mu_{N,t} (\cdot, \omega) \approx m_t, \ t \in [1,T] | \mu_{N,0} \approx m_0 \right] \asymp \exp \left( - N \calJ_{\rho,[0,T]} (m_{[1,T]} | m_0) \right).
\end{align}
Similar to the canonical setup, this assertion is assumed to be true for $\bQ_{\rho}$-almost every $\omega$. At the initial time, we directly verify the large deviation property,
\begin{align} \label{eq:mcild}
 U_N \left[ \mu_{N,0} \approx m_0 \right] \asymp \exp (N H(m_0)),
\end{align}
where
\begin{align}
 H(m) \coloneqq - \frac{1+m}{2} \ln \frac{1+m}{2} - \frac{1-m}{2} \ln \frac{1-m}{2} - \ln 2
\end{align}
is the Boltzmann entropy of the macroscopic state specified by the value of the magnetization $m \in [-1,1]$. We remark that the Boltzmann entropy increases in time along the typical path, i.e., $H ( \Phi_{\rho}^t (m)) > H(m)$ for $t \in \PZ$, which is regarded as the second law of thermodynamics in the relaxation process for the Kac ring model. Combining Eqs. (\ref{eq:mcld}) with (\ref{eq:mcild}), we have the large deviation property in equilibrium,
\begin{align}
 U_N \left[ \mu_{N,t} (\cdot, \omega) \approx m_t, \ t \in [0,T] \right] \asymp \exp \left( - N \left[ \calJ_{\rho,[0,T]} (m_{[1,T]} | m_0) - H(m_0) \right] \right).
\end{align}
From the definition of the canonical measure, Eq. (\ref{eq:cm}), we obtain 
\begin{align}
 \bP_{N;m} \left[ \mu_{N,0} \approx m_0 \right] \asymp \exp \left( - N ( - \beta m_0 + P(\beta) - H(m_0)) \right)
\end{align}
and
\begin{align}
 &\bP_{N;m} \left[ \mu_{N,t} \approx m_t, \ t \in [0,T] \right] 
 \notag \\
 &\asymp \exp \left( - N ( - \beta m_0 + P(\beta) + \calJ_{\rho,[0,T]}(m_{[1,T]} | m_0) - H(m_0) ) \right).
\end{align}
Hence, the rate function for the deviation probabilities conditioned on the initial magnetization in the canonical setup is no longer dependent on parameter $m$, and is given by the rate function $\calJ_{\rho}$ in the microcanonical setup,
\begin{align}
 \bP_{N;m} \left[ \mu_{N,t} (\cdot, \omega) \approx m_t, \ t \in [1,T] | \mu_{N,0} \approx m_0 \right] \asymp \exp \left( - N \calJ_{\rho,[0,T]} (m_{[1,T]} | m_0) \right)
\end{align}
for $\bQ_{\rho}$-almost every $\omega$. Moreover, we obtain that
\begin{align}
 \calJ_{\rho,[0,T]} (m_{[1,T]} | m_0 ) = \calI_{m,\rho,[0,T]}(m_{[0,T]}) - \calI_{m,\rho,0}(m_0) = \calI_{m_0,\rho,[0,T]}(m_{[0,T]}).
\end{align}
Therefore, the canonical measure biases the uniform measure so that the initial magnetization is fixed to a specific value $m$ and does not affect the rate function of the transition probabilities. We note that this property is consistent with the Onsager-Machlup formula (\ref{eq:om}) because there we consider small deviations $m_0 - m = O(N^{-1/2})$, and therefore $\calL_{m,\rho}^q = \calL_{m_0,\rho}^q + O(N^{-1/2})$.

\subsection{Detailed Fluctuation Relation}

We consider an important implication of microscopic reversibility of the Kac ring model. The argument below is based on Refs. \cite{MaesNetocnyShergelashvili2009,DeRoeckMaesNetocny2006}. The invariance of $U_N$ under the dynamics $\varphi_{N;\omega}$ implies that
\begin{align} \label{eq:reversibility}
 U_N \left[ m_N \circ \varphi_{N;\omega}^t \approx m_t, \ t \in [0,T] \right] = U_N \left[ m_N \circ \overline{\varphi}_{N;\omega}^t \approx m^*_t, \ t \in [0,T] \right],
\end{align}
where $m^*_t \coloneqq m_{T-t}$ for $t \in [0,T]$. We now introduce time reversal operators $\pi_N : S^N \to S^N$ and $\theta_N : B^{\PZ} \to B^{\PZ}$ defined as
\begin{align}
 (\pi_N \sigma_N)(i) = \begin{cases}
 \sigma_N (i) & i = 1 \\
 \sigma_N (i) & N \text{ is even and } i = \frac{N}{2} + 1 \\
 \sigma_N (N - i + 2) & \text{otherwise}
 \end{cases}
\end{align}
and
\begin{align}
 (\theta_N \omega)(i) = \begin{cases}
 \omega (N-i+1) & i = 1, \dots, N \\
 \omega (i) & \text{otherwise}.
 \end{cases}
\end{align}
Then, we express microscopic reversibility as $\pi_N \circ \varphi_{N; \theta_N \omega} \circ \pi_N = \overline{\varphi}_{N; \omega}$. This expression is easily understood noting that maps $\varphi_{N;\omega}$ and $\overline{\varphi}_{N;\omega}$ correspond, respectively, to the right-handed and left-handed rotations of the ring (Fig.\ref{fig:kacring}).The invariance of the uniform measure and the magnetization under the time reversal are easily verified; specifically, $U_N \pi_N^{-1} = U_N$ and $m_N \circ \pi_N = m_N$. Using microscopic reversibility and these invariances, along with Eq. (\ref{eq:reversibility}), we have that
\begin{align} \label{eq:reversibility2}
 U_N \left[ m_N \circ \varphi_{N;\omega}^t \approx m_t, \ t \in [0,T] \right] = U_N \left[ m_N \circ \varphi_{N; \theta_N \omega}^t \approx m^*_t, \ t \in [0,T] \right].
\end{align}
We recall that the time reversal of the disorder $\omega \mapsto \theta_N \omega$ inverts spatially the configuration of the scatterers and does not change the density of scatterers. We then expect that
\begin{align}
 U_N \left[ m_N \circ \varphi_{N; \theta_N \omega}^t \approx m^*_t, \ t \in [0,T] \right] \asymp \exp \left( - N \left[ \calJ_{\rho,[0,T]} (m^*_{[0,T]} | m_0^*) - H(m^*_0) \right] \right)
\end{align}
for $\bQ_{\rho}$-almost every $\omega$. With this expectation, we obtain from Eq. (\ref{eq:reversibility2}) that
\begin{align} \label{eq:fr}
 H(m_T) - H(m_0) = \calJ_{\rho,[0,T]}(m^*_{[0,T]} | m_T) - \calJ_{\rho,[0,T]} (m_{[0,T]} | m_0).
\end{align}
This relation is valid for generic systems having macroscopic autonomy \cite{DeRoeckMaesNetocny2006}. The left-hand side corresponds to the production of the Boltzmann entropy along the macroscopic path $m_{[0,T]}$. Therefore, the equality (\ref{eq:fr}) is regarded as a \textit{detailed fluctuation relation}. In particular, Eq. (\ref{eq:fr}) and the property $\calJ_{\rho,[0,T]}((\Phi_{\rho}^t (m_0))_{t \in [0,T]}|m_0) = 0$ lead to
\begin{align}
 \calJ_{\rho,[0,T]}( ( \Phi_{\rho}^{T-t}(m_0) )_{t \in [0,T]} | \Phi_{\rho}^T(m_0) ) = H ( \Phi_{\rho}^T (m_0)) - H(m_0) > 0 \ \text{for} \ T > 0,
\end{align}
which means that the probability that the time reversal of the most probable macroscopic path is realized is exponentially small with respect to system size $N$ and the rate is given by the Boltzmann entropy production along the typical path.

\subsection{A Direct Proof of Detailed Fluctuation Relation}

The fluctuation relation (\ref{eq:fr}) is expressed as the fluctuation symmetry of the quenched rate function in the canonical setup:
\begin{align} \label{eq:cfr}
 \calI_{m,\rho,[0,T]}^q (m^*_{[0,T]}) - \calI_{m,\rho,[0,T]}^q (m_{[0,T]}) = \beta (m_0 - m_T) = \frac{m_0 - m_T}{2} \ln \frac{1+m}{1-m}.
\end{align}
Equivalently, this symmetry is expressible in terms of the quenched generating function. By inserting the probability of the reversed path, we obtain that
\begin{align}
 &\Lambda_{m,\rho,[0,T]}^q (k_0, k_1, \dots, k_T) 
 \notag \\
 &= \lim_{N \to \infty} \frac{1}{N} \ln \sum_{m_{[0,T]}} \bP_{N;m} ( \mu_{N,t} (\cdot, \omega) \approx m_t, \ t \in [0,T] ) \exp \left( N \sum_{t=0}^T k_t m_t \right)
 \notag \\
 &= \lim_{N \to \infty} \frac{1}{N} \ln \sum_{m_{[0,T]}} \frac{\bP_{N;m} ( \mu_{N,t} (\cdot, \omega) \approx m_t, \ t \in [0,T] )}{\bP_{N;m} (\mu_{N,t} (\cdot, \omega) \approx m^*_t, \ t \in [0,T] )}
 \notag \\
 & \qquad \times \bP_{N;m} (\mu_{N,t} (\cdot, \omega) \approx m^*_t, \ t \in [0,T] ) \exp \left( N \sum_{t=0}^T k_{T-t} m^*_t \right).
\end{align}
The fluctuation relation (\ref{eq:cfr}) implies that
\begin{align}
 \frac{\bP_{N;m} ( \mu_{N,t} (\cdot, \omega) \approx m_t, \ t \in [0,T] )}{\bP_{N;m} (\mu_{N,t} (\cdot, \omega) \approx m^*_t, \ t \in [0,T] )} \asymp \exp \left( N \beta (m_0 - m_T) \right).
\end{align}
Therefore, we have the fluctuation relation in terms of the generating function,
\begin{align} \label{eq:cgfr}
 \Lambda_{m,\rho,[0,T]}^q (k_0, k_1, \dots, k_T) = \Lambda_{m,\rho,[0,T]}^q (k_T - \beta, k_{T-1}, \dots, k_1, k_0 + \beta )
\end{align}
with $\tanh \beta = m$. We prove the relation (\ref{eq:cgfr}) directly based on the form of the generating function (\ref{eq:qgf}). For $\eta_1, \dots, \eta_T \in S$, we find that
\begin{align*}
 & \cosh \left( k_T - \beta + \sum_{t=1}^{T-1} k_{T-t} \eta_{T-t+1} \dots \eta_{T} + (k_0 + \beta) \eta_T \dots \eta_1 \right) 
 \notag \\
 &= \cosh ( \alpha + \beta ( 1- \eta_1 \dots \eta_T)),
 \notag \\
 & \sinh \left( k_T - \beta + \sum_{t=1}^{T-1} k_{T-t} \eta_{T-t+1} \dots \eta_{T} + (k_0 + \beta) \eta_T \dots \eta_1 \right)
 \notag \\
 &= \eta_1 \dots \eta_T \sinh ( \alpha + \beta ( 1- \eta_1 \dots \eta_T)),
\end{align*}
where $\alpha = k_0 + \sum_{t=1}^{T} k_t \eta_1 \dots \eta_t$. It is easy to see that for either $\eta_1 \dots \eta_T =1$ or $\eta_1 \dots \eta_T = -1$, 
\begin{align*}
 &\cosh ( \alpha + \beta ( 1- \eta_1 \dots \eta_T)) + \tanh \beta \sinh ( \alpha + \beta ( 1- \eta_1 \dots \eta_T))
 \notag \\
 &=  \cosh \alpha + \tanh \beta \sinh \alpha.
\end{align*}
Hence, we obtain Eq. (\ref{eq:cgfr}) from the above relation and Eq. (\ref{eq:qgf}).

\section{Conclusions}
\label{sec:conclusion}

As an illustrative example of macroscopic dynamical fluctuations in deterministic systems, we studied fluctuations in the Kac ring model. Specifically, we derived the form of the generating function of the Kac ring model in the quenched setting. From this result, we have found that small deviations around the most probable path are described by the discrete-time Ornstein-Uhlenbeck process, and have obtained the Onsager-Machlup form of the path probability. Furthermore, based on the form of the generating function, we proved the fluctuation symmetry of the rate function originated from the microscopic reversibility.

In concluding, we comment on the future directions of this study. First, an interesting topic to investigate is the large deviation property of other models such as high-dimensional extension \cite{Lefevere2013} and quantum extension \cite{DeRoeckJacobsMaesNetocny2003} of the Kac ring model. In the high-dimensional ring model, the total number of particles on the rings is conserved and the density profile obeys a discrete-time and discrete-space diffusion equation. The study of the fluctuating diffusion equation for this model may provide a new insights into macroscopic fluctuation theory \cite{BertiniDeSoleGabrielliJonaLasinioLandim2007}, which was mainly developed for stochastic systems.

Second, it would be interesting to study dynamical fluctuations in the scaling limit. Ref. \cite{GottwaldOliver2009} considered the typical macroscopic law in the scaling limit of the Kac ring model. Since any phase points are periodic with the periods at most $2N$, it is reasonable to employ the rescaling of time, $\tau = t N^{-\alpha}$, for $\alpha \in (0,1)$. To obtain the non-trivial macroscopic law in this scaling, we have to take $\rho = r N^{- \alpha} / 2$ for $r > 0$ so that the number of scatterers within $N^{\alpha}$ sites over which the spins pass in $\tau \sim 1$ remains finite. Then, the macroscopic law becomes $\Phi_{\rho}^t (m) \to \exp (- r \tau) m$ in the scaling limit and it is easy to prove the corresponding law of large numbers. It is natural to ask how the dynamical fluctuations looks like in this scaling limit. For instance, we can see that in this scaling limit, Eq. (\ref{eq:qOU}) becomes formally an Ornstein-Uhlenbeck process,
\begin{align}
 d\nu_{\tau} = - r \nu_{\tau} d \tau + \sqrt{2r (1-m^2)} dW_{\tau},
\end{align}
with the initial data $\nu_0 \sim \calN (0,1-m^2)$, where $W_{\tau}$ is the Wiener process. The detailed analysis on dynamical fluctuations in the scaling limit is left for future work.

\begin{acknowledgements}
The author thanks Shin-ichi Sasa for making useful comments. The author especially thanks the anonymous referee for valuable comments that improved this paper. The present work was supported by JSPS KAKENHI Grant Number JP20J12143.
\end{acknowledgements}

\appendix

\section{Explicit Forms of Quenched Rate Function for $T=0$ and $T=1$}
\label{app:explicit}

We present the explicit form of the quenched rate function for $T=0$ and $T=1$. Since the initial distribution is the Bernoulli distribution, the quenched and annealed rate function at $T=0$ is given by
\begin{align}
 &\calI_{m,\rho,0}^{q} (m_0)= \calI_{m,\rho,0}^{a} (m_0) = D \left( \frac{1+m_0}{2}, \frac{1-m_0}{2} \| \frac{1+m}{2}, \frac{1-m}{2} \right),
\end{align}
where $D( \{ p_i \} \| \{ q_i \} ) = \sum_{i} p_i \ln (p_i / q_i)$ denotes the Kullback-Leibler divergence between two probability densities $\{ p_i \}$ and $\{ q_i \}$. The quenched generating function in time interval $[0,1]$ is given by
\begin{align*}
 &\Lambda_{m,\rho,[0,1]}^q (k_0,k_1) 
 \notag \\
 &= \rho \ln (\cosh (k_0 - k_1) + m \sinh (k_0 - k_1)) + (1-\rho) \ln ( \cosh (k_0 + k_1) + m \sinh (k_0 + k_1)).
\end{align*}
In this case, we can calculate the Legendre transform (\ref{eq:lt}) explicitly,
\begin{align} \label{eq:rf01}
 \calI_{m, \rho,[0,1]}^q(m_0,m_1) &= 
 \begin{cases}
 \calI_{m,\rho,0}^q(m_0) + \calJ_{\rho,[0,1]}(m_1 | m_0) & \text{if} \ |m_0 - m_1| < 2 \rho \\ & \quad \land \ |m_0+m_1| < 2(1-\rho) \\
 \infty & \text{otherwise}
 \end{cases}
\end{align}
where $\calJ_{\rho,[0,1]}(m_1|m_0)$ is given by
\begin{align}
 &\calJ_{\rho,[0,1]}(m_1|m_0) 
 \notag \\
 &= h \left( \frac{1+m}{2}, \frac{1-m}{2} \right) + h ( \rho, 1 - \rho) 
 \notag \\
 & \ \ \ - h \left( \frac{2 \rho - m_0 + m_1}{4}, \frac{2 \rho + m_0 - m_1}{4}, \frac{2(1-\rho) + m_0 + m_1}{4}, \frac{2(1-\rho) - m_0 - m_1}{4} \right).
\end{align}
Here, $h ( \{ p_i \} ) = - \sum_i p_i \ln p_i$ denotes the Shannon entropy of the probability density $\{ p_i \}$. We note that $\calJ_{\rho}(m_1|m_0)$ is identical to the mutual information of the probability distribution $p_{\sigma, \omega} = [ 2 \rho^{\omega}(1-\rho)^{1 - \omega} + \sigma (m_0 + (1-2\omega) m_1)]/4$ on $S \times B$.

\end{document}